\documentstyle[a4]{article}

\input{psfig}
\def\Journal#1#2#3#4{#1 {\bf #2}, #3 (#4)}
\def\Book#1#2#3{{\it #1}, (#2, #3)}

\def\be{\begin{equation}}
\def\ee{\end{equation}}
\def\ba{\begin{array}}
\def\ea{\end{array}}
\def\bea{\begin{eqnarray}}
\def\eea{\end{eqnarray}}
\def\ds{\displaystyle}

\def\id{\mbox{\bf 1}}
\def\hf{\left(\frac{+1}{2!}\right)}
\def\nn{\nonumber}
\def\re{\mbox{\sf Re}}
\def\im{\mbox{\sf Im}}
\def\r{\mbox{\sf R}}
\def\z{\mbox{\sf Z}}
\def\c{\mbox{\sf C}}
\def\sr{\mbox{\scriptsize\sf R}}

\def\sc{\mbox{\scriptsize\sf C}}

\begin{document}

\thispagestyle{empty}
\setcounter{page}{0}
\vspace*{.6truein}

\centerline{\Large \bf Analytic Continuation of Bernoulli Numbers,}
\vskip .1truein
\centerline{\Large \bf a New Formula for the Riemann Zeta Function,}
\vskip .1truein
\centerline{{\Large \bf and the Phenomenon of Scattering of Zeros}
\footnote{Preprint DAMTP-R-97/19 on-line at 
http://www.damtp.cam.ac.uk/user/scw21/papers/}}

\vskip .5truein
\centerline{S.C. Woon}
\vskip .2truein
\centerline{Department of Applied Mathematics and Theoretical Physics}
\centerline{University of Cambridge, Silver Street, Cambridge CB3 9EW,
  UK}
\vskip .1truein
\centerline{Email: S.C.Woon@damtp.cam.ac.uk}
\vskip .3truein

\centerline{\bf Abstract}

The method analytic continuation of operators acting integer $n$-times
to complex $s$-times (hep-th/9707206) is applied to an operator that
generates Bernoulli numbers $B_n$ (Math. Mag., {\bf 70(1)}, 51 (1997)).
$B_n$ and Bernoulli polynomials $B_n(s)$ are analytic continued to
$B(s)$ and $B_s(z)$. A new formula for the Riemann zeta function
$\zeta(s)$ in terms of nested series of $\zeta(n)$ is derived. The new
concept of dynamics of the zeros of analytic continued polynomials is
introduced, and an interesting phenonmenon of `scatterings' of the
zeros of $B_s(z)$ is observed.

\vfil \eject

\section{Introduction: Bernoulli Numbers}

Bernoulli numbers $B_n$ were discovered by Jakob Bernoulli
(1654-1705) \cite{bernoulli}. They are defined \cite{bateman}
\cite{abramowitz} as
\be \frac{z}{e^z-1} \,=\, \sum_{n=0}^\infty \frac{B_n}{n!} z^n\;,
\quad |z| < 2\pi\;, \;\; n = 1,2,3,\dots \in \mbox{\z}^+
\label{def_B_n} \ee
Expanding the l.h.s. as a series and matching the coefficients on both
sides gives
\be B_1 = -1/2, \quad B_n \left\{ \ba{ccc} = 0 &,& \mbox{odd }n\\ \ne 0 &,&
    \mbox{even }n \ea \right. \ee
With this result, (\ref{def_B_n}) can be rewritten as
\be \frac{z}{e^z-1} + \frac{z}{2} \,=\, \sum_{n=0}^\infty
\frac{B_{2n}}{2n!} z^{2n} \label{def_B_2n} \ee
Alternatively, Bernoulli numbers can be defined as satisfying the
recurrence relation
\be B_n \,=\, -\, \frac{1}{n+1} \,\sum_{k=0}^{n-1}\, {n+1 \choose k}\,
B_k\;, \quad B_0 = 1 \label{def_B_n_recursion} \ee

Bernoulli numbers are interesting numbers. They appear in connection
with a wide variety of areas, from Euler-Maclaurin Summation formula
in Analysis \cite{woon_tree} \cite{spivak} and the Riemann zeta
function in Number Theory \cite{riemann_zeta_function}
\cite{woon_period_jump}, to Kummer's regular primes in special cases of
Fermat's Last Theorem and Combinatorics \cite{kummer}.

\section{A Tree for Generating Bernoulli Numbers}

It was shown in \cite{woon_tree} how a binary Tree for generating
Bernoulli numbers can be intuited step-by-step and eventually
discovered. In the process of calculating the analytic continuation
of the Riemann zeta function to the negative half plane term-by-term,
an emerging pattern was observed.  The big picture of the structure of
the Tree became apparent on comparing the derived expressions with the
Euler-Maclaurin Summation formula.

In this paper, we start with the Tree and proceed on to find
interesting applications. While doing so, we will encounter some
surprising consequences.

The Tree can be constructed using two operators, $O_L$ and $O_R$.

At each node of the Tree sits a formal expression of the form $\ds
\frac{\pm 1}{a!\,b!\dots}$.

Define $O_L$ and $O_R$ to act only on formal expressions of this
form at the nodes of the Tree as follows:
\bea O_L & : & \frac{\pm 1}{a!\,b!\dots} \to 
 \frac{\mp 1}{(a+1)!\,b!\dots} \\
 O_R & : & \frac{\pm 1}{a!\,b!\dots} \to 
 \frac{\pm 1}{2!\,a!\,b!\dots} \eea

Schematically, \begin{itemize}
\item $O_L$ acting on a node of the Tree generates a branch downwards to
the left (hence the subscript ${}_L$ in $O_L$) with a new node at the
end of the branch.
\item $O_R$ acting on the same node generates a branch downwards to the
right. \end{itemize}
\begin{figure}[hbt]
\begin{center}
\begin{tabular}{c}
\psfig{figure=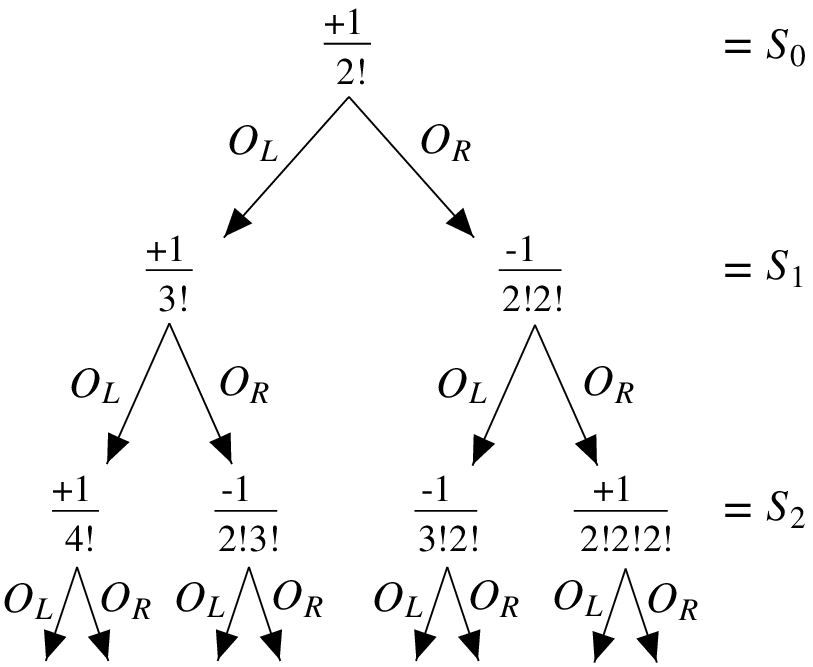,height=160pt}
\end{tabular}
\caption{The binary Tree that generates Bernoulli numbers.}
\end{center}
\end{figure}
Form a finite series out of the sum of the two non-commuting operators
\be S_n \,=\, (O_L + O_R)^n \hf \,=\, \left( O_L^n + \sum_{k=0}^{n-1} 
O_L^{n-1-k} O_R O_L^k + \cdots + O_R^n \right) \hf \ee
This is equivalent to summing terms on the $n$-th row of nodes across
the Tree.

Bernoulli numbers are then simply given by
\be B_n \,=\, n! \; S_{n-1} \quad \mbox{for } \;n \ge 2 \ee
$$ \mbox{eg., } B_3 \,=\, 3!\; S_2 \,=\, 3!\; (O_L + O_R)^2
\left(\frac{+1}{2!}\right) \,=\, 3!\;(O_L + O_R)\,(O_L+O_R)
\left(\frac{+1}{2!}\right)$$ 
$$ =\, 3!\;(O_L O_L + O_L O_R + O_R O_L + O_R O_R) \left(\frac{+1}{2!}\right)$$
$$=\, 3!\;\left( \frac{+1}{4!} + \frac{-1}{2!3!} + \frac{-1}{3!2!} +
\frac{+1}{2!2!2!} \right) = 0$$

By observation, this Sum-across-the-Tree representation of $\,S_n\,$
is exactly equivalent to the following determinant known to generate $B_n$,
\be S_n = (-1)^n \left| \ba{ccccccc}
\ds\Bigg.\frac{1}{2!}&1&0&\;\quad 0\quad\;&\;\quad 0\quad\;&
 \;\cdots\;&\;\;\;0\;\\
\ds\Bigg.\frac{1}{3!}&\ds\frac{1}{2!}&1&0&0&\cdots&\;\;\;0\;\\
\ds\Bigg.\frac{1}{4!}&\ds\frac{1}{3!}&\ds\frac{1}{2!}&1&0&\cdots&\;\;\;0\;\\
\ds\Bigg.\vdots&\ddots&\ddots&\ddots&\ddots&\ddots&\;\;\;\vdots\;\\
\ds\Bigg.\frac{1}{(n-2)!}&\ddots&\ddots&\ds\frac{1}{3!}&\ds\frac{1}{2!}&1&
 \;\;\;0\;\\
\ds\Bigg.\frac{1}{(n-1)!}&\ds\frac{1}{(n-2)!}&\ddots&\ddots&
 \ds\frac{1}{3!}&\ds\frac{1}{2!}&\;\;\;1\;\\
\ds\Bigg.\frac{1}{n!}&\ds\frac{1}{(n-1)!}&\ds\frac{1}{(n-2)!}&
 \ddots&\ddots&\ds\frac{1}{3!}&\;\;\;\ds\frac{1}{2!}\;
\ea \right| \ee

\section{Analytic Continuations}
\subsection{Analytic Continuation of Operator}

First, we introduce the idea of analytic continuing the action of an
operator following \cite{woon_operator}. We are used to thinking of an
operator acting once, twice, three times, and so on. However, an
operator acting integer times can be analytic continued to an operator
acting complex times by making the following observation:

A generic operator $A$ acting complex $s$-times can be formally
expanded into a series as
\bea 
  A^s&=&\left( \Big. w \id - \left[ \big. w \id - A \right]
   \right)^{\!s} = w^s\Big(1-\Big[1-
\frac{1}{w} A\Big]\Big)^{\!s}\nn\\ 
&=& w^s\!\left( \id + \sum_{n=1}^\infty \frac{(-1)^n}{n!} \left[
    \prod_{k=0}^{n-1} (s \!-\! k) \right]\!\left[ \big. \id - \frac{1}{w}A
  \right]^{\!n} \right) \nn\\ 
&=& w^s\!\left(\id + \sum_{n=1}^\infty \frac{(-1)^n}{n!} \left[
    \prod_{k=0}^{n-1} (s \!-\! k) \right] \! \left[ \id + \sum_{m=1}^n 
     \left(\frac{-1}{w}\right)^{\!\!m} \!\! {n \choose m} A^m \right]
      \right)
\label{op_series}
\eea where $\;s\in\c$, $w\in\r$, and $\;\id\;$ is
the identity operator.

The region of convergence in $\,s\,$ and the rate of convergence of the
series will in general be dependent on operator $A,\;$ parameter
$w,\;$ and the operand on which $A$ acts.
 
\subsection{Analytic Continuation of the Tree-Generating Operator}

Just as in (\ref{op_series}), the Tree-generating operator $(O_L
+ O_R)$ acting $(s-1)$ times on $\ds\hf$ can be similarly expanded as
\bea \lefteqn{(O_L + O_R)^{s-1} \hf}\nn\\
\; &=& w^s\left(\id + \sum_{n=1}^\infty \frac{(-1)^n}{n!} \left[
    \prod_{k=1}^n (s \!-\! k) \right] \! \left[ \id + \sum_{m=1}^n 
     \left(\frac{-1}{w}\right)^{\!\!m} \!\! {n \choose m} (O_L + O_R)^m \right]
      \right) \!\hf \nn\\ 
\; &=& w^s\left(\frac{1}{2} + \sum_{n=1}^\infty \frac{(-1)^n}{n!}
      \left[ \prod_{k=1}^n (s \!-\! k) \right] \! \left[
        \frac{1}{2} + \sum_{m=1}^n \left(\frac{-1}{w}\right)^{\!\!m} \!\! {n
          \choose m} \frac{B_{m+1}}{(m\!+\!1)!}\right]\right)
 \label{op_tree_series}\eea
which converges for \re$(s) > (1/w)\;$ where $\;s\in\c$, $\;w\in\r$,
 $\;w > 0$.

\subsection{Analytic Continuation of Bernoulli Numbers}

\be B_n \,=\, n! \; (O_L + O_R)^{n-1} \hf \,=\, \Gamma(1+n) \; 
(O_L + O_R)^{n-1} \hf\nn\ee

Now that we can analytic continue the Tree-generating operator $(O_L +
O_R)$ with (\ref{op_tree_series}), if we do so, we turn the sequence of
Bernoulli numbers $B_n$ into their analytic continuation --- a function
\bea \lefteqn{B(s) \;=\; \Gamma(1+s) \; (O_L + O_R)^{s-1} \hf}
 \label{B_series}\\
\;&=& w^s\;\Gamma(1+s) \left(
 \frac{1}{2} + \sum_{n=1}^\infty \frac{(-1)^n}{n!}
      \left[ \prod_{k=1}^n (s \!-\! k) \right] \! \left[
        \frac{1}{2} + \sum_{m=1}^n \left(\frac{-1}{w}\right)^{\!\!m} \!\! {n
          \choose m} \frac{B_{m+1}}{(m\!+\!1)!} \right] \right)\nn\eea
which converges for $\;$\re$(s) > (1/w)\;,\;$ real $\;w > 0.$

So effectively, by the method of analytic continuation of operator, we
have now obtained the function $\;B(s)\;$ as the analytic continuation of
Bernoulli numbers.
\begin{figure}[hbt]
\begin{center}
\begin{tabular}{c}
\psfig{figure=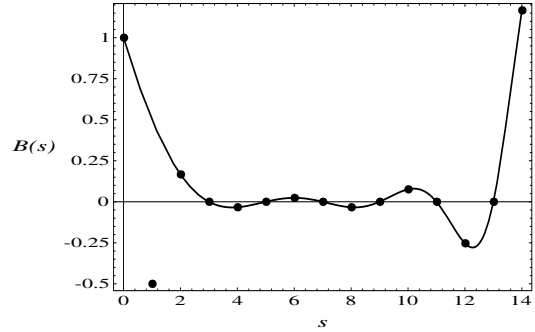,height=125pt,width=200pt}
\end{tabular}
\caption{The curve $\,B(s)\,$ runs through the points of all $\,B_{n}$
  except $\,B_1$.}
\end{center}
\end{figure}

All the Bernoulli numbers $B_n$ agree with $B(n)$, the analytic
continuation of Bernoulli numbers evaluated at $n$,
\be B(n) \,=\, B_n \quad\mbox{for}\quad n \ge 2\ee
\be\mbox{except} \quad\quad B(1) = \frac{1}{2} \quad \mbox{
  but } \quad B_1 = -\,\frac{1}{2} \quad\quad\quad\quad
\label{B_1}\ee

\section{Missing Signs in the Definition of $B_n$}

Looking back at (\ref{def_B_n}) to (\ref{def_B_n_recursion}), we can
see that the sign convention of $B_1$ was actually arbitrary.
(\ref{B_1}) suggests that consistent definition of Bernoulli numbers
{\em should really have been}
\be \frac{z}{e^z-1} \,=\, \sum_{n=0}^\infty (-1)^n
\frac{B_n}{n!}  z^n\;, \quad |z| < 2\pi\;, \;\; n = 1,2,3,\dots \in
\mbox{\z}^+ \label{newdef_B_n} \ee
or
\be B_n \,=\, \frac{(-1)^{n+1}}{n+1} \,\sum_{k=0}^{n-1}\, (-1)^k\, 
{n+1 \choose k}\, B_k\;, \quad B_0 = 1
\label{newdef_B_n_recursion} \ee
which only changes the sign in the conventional definition of the only
non-zero odd Bernoulli numbers, $B_1$, from $\;B_1 = -1/2\;$ to 
$\;B_1 = B(1) = 1/2\;$.
 
So here's my little appeal to the Mathematics, Physics, Engineering,
and Computing communities to introduce the missing signs into
the sum in the definition of Bernoulli numbers as in
(\ref{newdef_B_n}) and (\ref{newdef_B_n_recursion}) because the
analytic continuation of Bernoulli numbers fixes the arbitrariness of
the sign convention of $B_1$.

\section{A New Formula for the Riemann Zeta Function}

Bernoulli numbers are related to the Riemann zeta function as 
\cite{bateman} \cite{abramowitz}
\be \zeta(-n) \,=\, - \;\frac{B_{n+1}}{n+1} \ee
\be \zeta(2n) \,=\, \frac{(-1)^{n+1}(2\pi)^{2n}}{2\;(2n)!} B_{2n}
\label{zeta(2n)}\ee 
for $\;n = 0,1,2,\dots \in$ \z${}^+_0$.

From the above analytic continuation of Bernoulli numbers,
\bea B_n &\mapsto& B(s) \nn \\
\left( \zeta(-n) \,=\, - \;\frac{B_{n+1}}{n+1} \right) &\mapsto&
\left( \zeta(-s) \,=\, - \;\frac{B(s+1)}{s+1} \right) \nn \\
\Rightarrow \quad \zeta(1-s) &=& - \;\frac{B(s)}{s} \label{zeta(1-s)}\eea
Replacing $B(s)$ in (\ref{zeta(1-s)}) with the series in
(\ref{B_series}) and noting that $\;\Gamma(1+s)/s = \Gamma(s)\;$ gives
\be \zeta(1-s) = -\;w^s\;\Gamma(s) \left( \frac{1}{2} + \sum_{n=1}^\infty
  \frac{(-1)^n}{n!}  \left[ \prod_{k=1}^n (s \!-\! k) \right] \!
  \left[ \frac{1}{2} + \sum_{m=1}^n \left(\frac{-1}{w}\right)^{\!\!m} \!\!
    {n \choose m} \frac{B_{m+1}}{(m+1)!} \right] \right)
\label{zeta(1-s)_series} \ee
which converges for \re$(1-s) < 1-(1/w)\;,\;$ real $\; w > 0$.

The functional equation of the Riemann zeta function relates
$\zeta(1-s)$ to $\zeta(s)$ as
\be \zeta(1-s) = 2 \;(2\pi)^{-s} \;\Gamma(s)\; \cos\left(\frac{\pi
    s}{2}\right) \;\zeta(s) \ee

Applying this relation to (\ref{zeta(1-s)_series}) yields
\be \!\!\!\!\!\!\!\!\!\!\! \cos\left(\frac{\pi \hat{s}}{2}\right)
\zeta(s) = -\; \frac{(2\pi w)^s}{2} \left( \frac{1}{2} +
  \sum_{n=1}^\infty \frac{(-1)^n}{n!}  \left[ \prod_{k=1}^n (s \!-\!
    k) \right] \!  \left[ \frac{1}{2} + \sum_{m=1}^n
    \left(\frac{-1}{w}\right)^{\!\!m} \!\!  {n \choose m}
    \frac{B_{m+1}}{(m+1)!} \right] \right) \ee
or in the limiting form
$$ \!\!\!\!\!\!\!\!\!\!\zeta(s) = -\; \frac{(2\pi w)^s}{2} 
  \lim_{\;{\ds \hat{s}}\to {\ds s}} \left\{\!
  \frac{ \left( \ds \frac{1}{2} + \sum_{n=1}^\infty \frac{(-1)^n}{n!}
  \left[ \prod_{k=1}^n (\hat{s} \!-\! k) \right] \left[ \frac{1}{2} +
  \sum_{m=1}^n \left(\frac{-1}{w}\right)^{\!\!m} \!\!  {n \choose m}
  \frac{B_{m+1}}{(m+1)!} \right] \!\right)}
{\ds \cos\left(\frac{\pi \hat{s}}{2}\right)} \right\} $$
\be = -\; \frac{(2\pi w)^s}{2} 
  \lim_{\;{\ds \hat{s}}\to {\ds s}} \left\{\!
  \frac{ \left( \ds \frac{1}{2} + \sum_{n=1}^\infty \frac{(-1)^n}{n!}
  \left[ \prod_{k=1}^n (\hat{s} \!-\! k) \right] \left[ \frac{1}{2} -
  \sum_{m=1}^n \left(\frac{-1}{w}\right)^{\!\!m} \!\!  {n \choose m}
  \frac{\zeta(-m)}{m!} \right] \!\right)}
{\ds \cos\left(\frac{\pi \hat{s}}{2}\right)} \right\}
  \label{zeta(s)_zeta(-n)} \ee
a nested sum of the Riemann zeta function itself evaluated at negative
integers, which converges for \re$(s) > (1/w)\;,\;$ real $\; w > 0
\;,\;$ and the limit only needs to be taken when $\;s = 1,3,5,\dots
\in$ \z${}^+_{odd}$, the set of positive odd integers, for which the
denominator $\;\ds\cos\left(\frac{\pi s}{2}\right) = 0$.

This is consistent with 
$$ \left( \zeta(2n) \,=\,
  \frac{(-1)^{n+1}(2\pi)^{2n}}{2\;(2n)!} B_{2n} \right)
\;\;\mapsto\;\; \left( \zeta(2s) \,=\, -\, 
\frac{(2\pi)^{2s}}{\ds 2\; \Gamma(1+2s)}\,
\lim_{\;{\ds \hat{s}}\to {\ds s}}\,
\left\{\frac{B(2\hat{s})}{\cos(\pi \hat{s})} \right\}\right) $$
\be\Rightarrow \quad \zeta(s) \;\;=\;\; -\,
\frac{(2\pi)^{s}}{\ds 2\; \Gamma(1+s)}
\lim_{\;{\ds \hat{s}}\to {\ds s}}\, 
\left\{\frac{B(\hat{s})}{\cos\left(\ds\frac{\pi \hat{s}}{2}\right)}\right\} 
\quad \ee
\begin{figure}[hbt]
\begin{center}
\begin{tabular}{c}
\psfig{figure=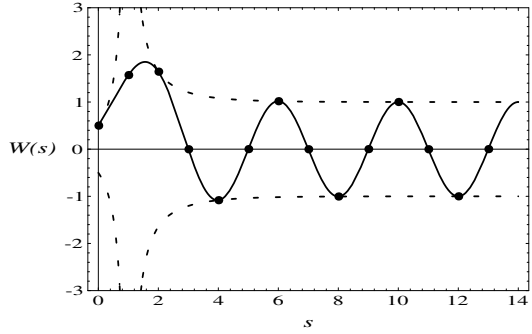,height=125pt,width=200pt}
\end{tabular}
\caption{$\ds \;W(s) \,=\, \frac{(2\pi)^{s} \; B(s)}{2 \;\Gamma(1+s)} = 
-\,\cos\left(\ds\frac{\pi s}{2}\right) \,\zeta(s).\;\; W(s)$ is bounded
by the envelopes $\Big.\,\pm \left|\zeta(s)\right|\;$ shown as dashed curves.}
\end{center}
\end{figure}

Since both $\,B(k)\,$ and $\,\cos\left(\ds\frac{\pi k}{2}\right) = 0\,$ 
for odd integer $\,k \in$ \z${}^+_{odd}\backslash\{1\}$, L'H\"opital rule
can be applied to the limit giving
\be \lim_{{\ds \hat{s}} \to {\ds s}} \, 
\left\{\frac{B(\hat{s})}{\cos\left(\ds\frac{\pi \hat{s}}{2}\right)}\right\} 
\,=\, \left\{ \ba{lcl}
\ds \frac{B(s)}{\cos\left(\ds\frac{\pi s}{2}\right)} &,& s \not\in
  \mbox{\z}^+_{odd}\\
& & \\
\infty &,& s = 1\\
& & \\
\ds (-1)^{(s+1)/2}\, \frac{2 B'(s)}{\pi} &,& 
s \in \mbox{\z}^+_{odd}\backslash\{1\} 
\ea \right. \ee
where the prime denotes differentiation and so $\,B'(s)\,$ is the derivative 
of the function $B(s)$.

Therefore we have now found the apparently missing `odd' expression
`dual' to the `even' expression (\ref{zeta(2n)}).
\be\mbox{(even):}\quad\quad\quad\quad\quad\quad\quad
 \zeta(2n) \,=\, (-1)^{n+1} \ds \frac{(2\pi)^{2n}}{(2n)!} \;
 \frac{B_{2n}}{2}\quad\quad\quad\quad\quad\quad\quad
\quad\quad\quad\;\ee
\be\mbox{(odd): }\quad\quad\quad\quad\quad
 \zeta(2n+1) \,=\, (-1)^n \ds \frac{(2\pi)^{2n+1}}{(2n+1)!} \;
 \frac{B'(2n+1)}{\pi}\quad\quad\quad\quad\quad\quad\quad\ee
for $\;n = 1,2,3,\dots \in$ \z$^+$.

In fact, we can express $\,B'(s)\,$ in terms of $\,\zeta'(s)$, the
derivative of $\,\zeta(s)$.
\bea
B(s) &=& -\; s\; \zeta(1-s)\nn\\
B'(s) &=& -\; \zeta(1-s) \,+\, s\; \zeta'(1-s)\nn\\
B'(2n+1) &=& \quad (2n+1)\; \zeta'(-2n)\\
& & \quad\quad\quad \mbox{as }\; \zeta(-2n) = 0\nn
\eea
\be \quad\quad\quad\quad\quad \Rightarrow \quad \quad
  \zeta(2n+1) \,=\, (-1)^n\, \frac{(2\pi)^{2n+1}}{(2n)!\;\pi}
  \;\zeta'(-2n) \;\quad \mbox{for } n \in \mbox{\z}^+.\ee
  
This is just the differential of the functional equation and so
verifies the consistency of $B(s)$ and $B'(s)$ with $B_n$ and $\zeta(s)$.

\section{Other Undiscovered Half of Bernoulli Numbers}
From the relation (\ref{zeta(1-s)}), we can define the other
analytic continued half of Bernoulli Numbers
$$ B(s) \,=\, -\;s\;\zeta(1-s)\;, \quad
B(-s) \,=\, s\;\zeta(s+1)$$
\be\Rightarrow\quad B_{-n} \,=\, B(-n) \,=\, n\;\zeta(n+1)\;, \quad n 
\in \mbox{\z}^+\ee
\indent Since $\;\zeta(n+1) \to 1\;$ as $\;n \to \infty, \;\; B_{-n} \sim n
\;$ asymptotically for $\; (-n) \ll -1$.
\begin{figure}[hbt]
\begin{center}
\begin{tabular}{c}
\psfig{figure=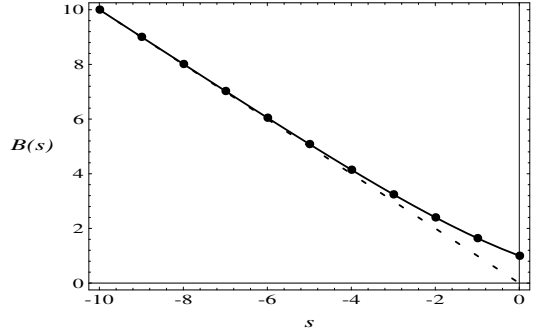,height=125pt,width=200pt}
\end{tabular}
\caption{The curve $\,B(s)\,$ runs through the points $\,B_{-n}$ and
  grows $\sim n\;$ asymptotically as $\,(-\,n)\, \to -\,\infty$.}
\end{center}
\end{figure}

\section{More Related Analytic Continuations}
\subsection{Analytic Continuation of Bernoulli Polynomials}
The conventional definition \cite{bateman} \cite{abramowitz} of
Bernoulli polynomials $\,B_n(x)\,$ also has an arbitrariness in the
sign convention.  For consistency with the redefinition of
$\,B_n\,=\,B(n)\,$ in (\ref{newdef_B_n}) and
(\ref{newdef_B_n_recursion}), Bernoulli polynomials should be
analogously {\em redefined} as
$$ \frac{z\;e^{xz}}{e^z-1} \,=\, \left( \sum_{k=0}^\infty (-1)^k B_k
  \frac{z^k}{r!} \right) \!\! \left( \sum_{m=0}^\infty
  \frac{(x\,z)^m}{m!} \right) \,=\, \sum_{n=0}^\infty (-1)^n B_n(x)
  \frac{z^n}{n!} \label{newdef_B_n(x)}$$
\be \Rightarrow \quad B_n(x) \,=\, \sum_{k=0}^n \,(-1)^{(n+k)}\, {n
    \choose k} \, B_k \, x^{n-k} \ee
The analytic continuation can be then obtained as
$$n \;\mapsto\; s\in\r\;,\;s \ge 1\;, \quad x \;\mapsto\; w\in\c$$
$$B_k \;\mapsto\; B\!\left(\big.k+(s\!-\![s])\right) \,=\, 
-\left(\big.k+(s\!-\![s])\right)\,\zeta\!\left(\big.1-(k+(s\!-\![s]))\right)$$
$${n \choose k} \;\mapsto\; \frac{\Gamma(1+s)}{\Gamma(1+k+(s\!-\![s]))\;
 \Gamma(1+[s]-k)}$$
$$\Rightarrow\quad B_n(x) \;\mapsto\; B(s,w) \,=\, \sum_{k=-1}^{[s]}
\frac{(1)^{n+k}\;\;\Gamma(1+s)\;\;B\!\left(\big.k+(s\!-\![s])\right)\;\;
w^{[s]-k}}{\Gamma(1+k+(s\!-\![s]))\;\Gamma(1+[s]-k)}$$
\be \quad\quad\quad\quad \sum_{k=0}^{[s]+1} \frac{(1)^{n+k+1}\;\;
\Gamma(1+s)\;\;B\!\left(\big.(k-1)+(s\!-\![s])\right)\;\;w^{[s]+1-k}}
{\Gamma(k+(s\!-\![s]))\;\Gamma(2+[s]-k)}\ee
where $\;[s]\;$ gives the integer part of $\;s\,,\;$ and so
$\;(s\!-\![s])\;$ gives the fractional part,
$$\mbox{eg.,}\quad B_2(x) \,=\, B(2,x) \,=\, 1/6 - x + x^2 \;\approx\,
0.16667 - x + x^2$$
$$ B(2.01,x) \,\approx\, 0.16420 - 0.99660\,x + 1.00920\,x^2 
 - 0.00554\,x^3 $$
$$ B(2.99,x) \,\approx\, 0.00092 - 0.50576\,x + 1.50001\,x^2 
 - 0.98744\,x^3$$
$$ B_3(x) \,=\, B(3,x) \,=\, - 0.5\,x + 1.5\,x^2 - x^3$$
\begin{figure}[hbt]
\begin{center}
\begin{tabular}{c}
\psfig{figure=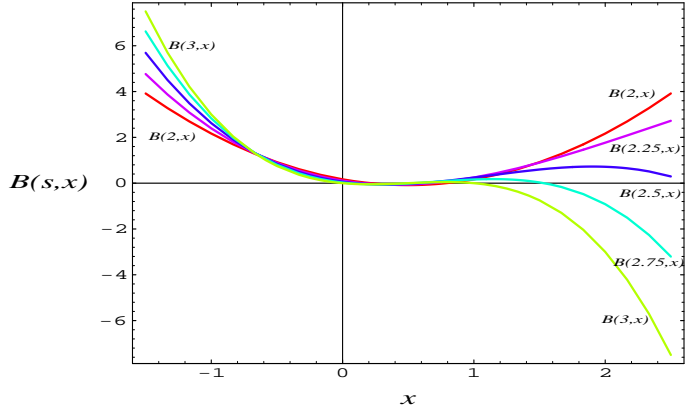,height=155pt,width=260pt}
\end{tabular}
\caption{Deformation of the curve of $\,B_2(x)\,=\,B(2,x)\,$ into the curve of
  $\,B_3(x)\,=\,B(3,x)\,$ via the real analytic continuation
  $\,B(s,x), \; 2 \le s \le 3, \; x\in\r\,.$}
\end{center}
\end{figure}

\subsection{Analytic Continuation of Euler Numbers and Euler Polynomials}

Euler numbers $\,E_n\,$ and Euler polynomials $\,E_n(x)\,$ are defined
\cite{bateman} \cite{abramowitz} as
\be \mbox{sech }z \,=\, \frac{2\;e^z}{e^{2z+1}} \,=\,
\sum_{n=0}^\infty E_n \frac{z^n}{n!}\,, \quad |z| < 2\pi\,,\;n \in
\mbox{\z}^+\;;\;\;\; E_n = 0 \;\;\mbox{for odd}\; n\ee
$$ \frac{2\;e^{xz}}{e^{2z+1}} \,=\, \frac{2\;e^{z/2}\;e^{z (x-1/2)}}
    {(e^z + 1)^{-1}} \,=\, \left( \sum_{k=0}^\infty E_k
     \frac{z^k}{2^k\,k!} \right) \!\! \left( \sum_{m=0}^\infty
      \frac{(x-\frac{1}{2}\Bigg.)^m z^m}{m!} \right) \,=\,
    \sum_{n=0}^\infty E_n(x) \frac{z^n}{n!} $$
\be \Rightarrow \quad E_n(x) \,=\, \sum_{k=0}^n {n \choose r} \,
    2^{-r} \, (x - \frac{1}{2})^{n-r} \, E_r \ee
and are related to Bernoulli polynomials as
\be E_n(x) \,=\, \frac{2}{n+1} \left( B_{n+1}(x) - 2^{n+1}
    B_{n+1}\!\left(\frac{x}{2}\right) \right) \ee
\be E_n \,=\, 2^n E_n\!\left(\frac{1}{2}\right) \,=\, \frac{2}{n+1} \left(
    B_{n+1}\!\left(\frac{1}{2}\right) - 2^{n+1}
    B_{n+1}\!\left(\frac{1}{4}\right) \right) \ee
Their analytic continuation then follows straightforwardly from the
analytic continuation of Bernoulli polynomials
$$ B_n(x) \mapsto B(s,w)\,,\quad E_n \mapsto E(s)\,,\quad
E_n(x) \mapsto E(s,w) $$
\centerline{where $\;n \mapsto s\in\r\;,\;s \ge 1\;, \quad x \mapsto
  w\in\c\,. \quad$}

\section{Beautiful zeros of Bernoulli Polynomials}

\subsection{Distribution and Structure of the zeros}

Zeros of Bernoulli polynomials are solutions of $\,B_n(w) = 0, \;w
\in$ \c.
\begin{figure}[hbt]
\begin{center}
\begin{tabular}{c}
\psfig{figure=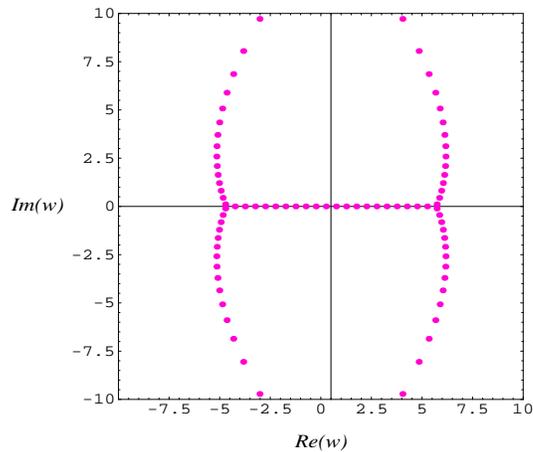,height=170pt,width=200pt}
\end{tabular}
\caption{Zeros of Bernoulli polynomials $B_{80}(w)$.}
\end{center}
\end{figure}
\begin{figure}[hbt]
\begin{center}
\begin{tabular}{c}
\psfig{figure=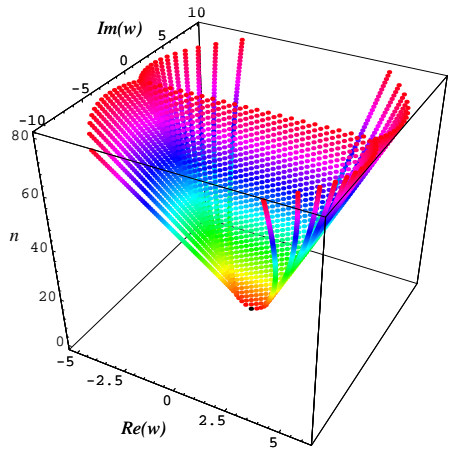,height=170pt}
\end{tabular}
\caption{Stacks of zeros of Bernoulli polynomials $B_n(w)$ form a
  3-D structure.}
\end{center}
\end{figure}

The real zeros, except the outermost pair in general, are
almost regularly spaced, while the complex zeros lie on arcs
symmetrical about \re$(w) = 1/2$.

\subsection{Observations, Theorem, Conjectures, and\\ Open Problems}

\begin{enumerate}
\item {\large \em Symmetries}
  
  Prove that $\,B_n(w),\; w\in\,\c,\,$ has $\,\re(w) = 1/2\,$
  reflection symmetry in addition to the usual $\,\im(w) = 0\,$
  reflection symmetry analytic complex functions. The obvious corollary is
  that the zeros of $\,B_n(w)\,$ will also inherit these symmetries.
  \be \mbox{If }\, B_n(w_0) = 0, \mbox{ then }\,
  B_n(1 - w_0) = 0 = B_n(w_0^\dag) = B_n(1 - w_0^\dag) \label{symmetries}\ee
  where ${}^\dag$ denotes complex conjugation.
  
\bigskip

\item{\large \em Non-degenerate zeros}

  Prove that $\,B_n(w) = 0\,$ has $\,n\,$ distinct solutions, ie., all the
  zeros are non-degenerate.

\bigskip

\item {\large \em Central zero Theorem}
  
  If $B_{2 m + 1}$ has $\,\re(w) = 1/2\,$ and $\,\im(w) = 0\,$
  reflection symmetries, and $(2 m + 1)$ non-degenerate zeros, then
  $(2 m)$ of the distinct zeros will satisfy (\ref{symmetries}). If
  the remaining one zero is to satisfy (\ref{symmetries}) too, it must
  reflect into itself, and therefore it must lie at 1/2, the center of
  the structure of the zeros, ie.,
  \be B_n(1/2) = 0 \quad \forall \; \mbox{ odd }\;n\ee

\item {\large \em Counting of real and complex zeros}
  
  Prove that the number of complex zeros {\bf z}${}_{\sc(n)}$ of
  $\,B_n(w)\,$ lying on the 4 sets of arcs off the real plane,
  $\,$\im$(w) \ne 0,\,$ is
  \be \mbox{\bf z}{}_{\sc(n)} = 4 \left[
    \frac{n-\delta}{5} \right]\,, \mbox{ where }\; \delta = \left[
    \frac{n+30}{21} \right]
  \label{interval} \ee
  and $\left[\big.\;\;\right]$ denotes taking the integer part. The
  factor 4 comes from the above 2 reflection symmetries.

\medskip

  Since $\,n\,$ is the degree of the polynomial $\,B_n(w),\,$ the
  number of real zeros {\bf z}${}_{\sr(n)}$ lying on the real plane \im$(w) =
  0\,$ is then $\;$ {\bf z}${}_{\sr(n)} = (n - ${\bf z}${}_{\sc(n)})$.

\medskip

See Appendix for tabulated values of {\bf z}${}_{\sr(n)}$ and {\bf
  z}${}_{\sc(n)}$.

\bigskip

\item {\large \em Asymptotic Lattice of real zeros}

  It is known that from (\ref{newdef_B_n(x)}),
  $$ \sum_{n=0}^{\infty} \left( B_n(w+1) - B_n(w) \right)
  \frac{z^n}{n!} \,=\, z\,e^{w z} \,=\, \sum_{n=1}^{\infty}
  \frac{w^{n-1}\,z^n}{(n-1)!} $$
  $$ \Rightarrow \quad B_n(w+1) - B_n(w) \,=\, w^{n-1} \quad\quad
  \quad\quad\quad\,$$
  $$ \Rightarrow \quad B_n(1) = B_n(0) = B_n = 0 \quad \forall
  \; \mbox{ odd } \, n \ge 3$$
  
  Show that all the real zeros of $\,B_n(w)\,$ except the outermost pair
  in general, are approximately regularly spaced at the staggered
  lattice points
  \be w = \left\{ \ba{lll}
  \Big. 0,\; \pm 1/2,\; \pm 1,\; \pm 3/2, \dots & \in \mbox{\z}/2 & 
  \mbox{ for odd }\,n\\
  \Big.\pm 1/4,\; \pm 3/4,\; \pm 5/4, \dots & \in \mbox{\z}/2 + 1/4 &
  \mbox{ for even }\,n \ea \right.\ee
  and becomes increasingly located exactly at these lattice points as
  $\,n \to \infty$.
\begin{figure}[hbt]
\begin{center}
$\begin{array}{c}
\psfig{figure=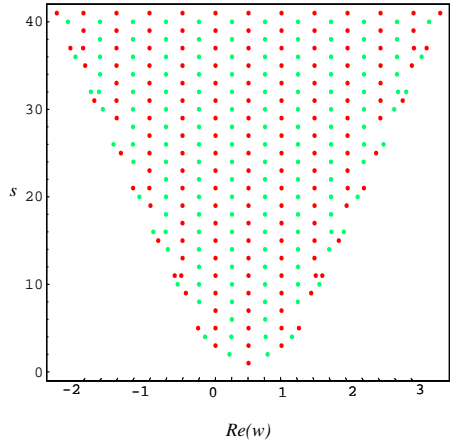,height=170pt}
\end{array}$
\caption{Inner real zeros converge to a staggered Lattice structure.}
\end{center}
\end{figure}

\smallskip
  See Appendix for tabulated solutions of $\,B_n(w) = 0\,.$

\bigskip

\item {\large \em Relation between zeros of Bernoulli and Euler
    polynomials}

  Choose any zero of Bernoulli polynomial $\,B_n(w)\,$ and denote it
  as $\, w_{B(n)}$,\\
  ie. $\,B_n(w_{B(n)}) = 0$.
  
\medskip

\noindent Prove that 
\be \mbox{as }\;n \to \infty\,,\;\;\; E_n\!\left(2\;( w_{B(n)} - 1/2 )
    + 1/2\right) = E_n\!\left(2\;w_{B(n)} - 1/2\right) = 0 \ee
  ie., the structure of the zeros of Euler polynomials resembles the
  structure of the zeros of Bernoulli polynomials but doubled in size
  in the limit the degree of the polynomials $\,n \to \infty$. Both
  structures are centered at $\,w = 1/2$.

\bigskip

\item {\large \em Bounding Envelopes and Trajectories of complex zeros}
  
  Find the equation of envelope curves bounding the real zeros lying
  on the plane, and the equation of a trajectory curve running
  through the complex zeros on any one of the arcs.
\end{enumerate}

\subsection{Dynamics of the zeros from Analytic Continuation}

Bernoulli polynomial $B_n(w)$ is a polynomial of degree $n$.  Thus,
$B_n(w)$ has $n$ zeros and $B_{n+1}(w)$ has $(n+1)$ zeros. When discrete
$n$ is analytic continued to continuous parameter $s$, it naturally
leads to the question:

How does $B(s,w)$, the analytic continuation of $B_n(w)$, pick up an
additional zero as $s$ increases continuously by one?

This introduces the exciting concept of the dynamics of the zeros of
analytic continued polynomials --- the idea of looking at how the
zeros move about in the $\,w\,$ complex plane as we vary the parameter
$\,s\,$.

Continuity shows that the additional zero simply ``flows in from
infinity''. 

To have a physical picture of the motion of the zeros in
the complex $w$ plane, imagine that each time as $s$ increases
gradually and continuously by one, an additional real zero flies in
from positive infinity along the real positive axis, gradually slowing
down as if ``it is flying through a viscous medium''.

For $s < 5$, the additional zero simply joins onto the almost
regularly spaced array of the real zeros streaming slowly towards the
negative real direction. The array of zeros continue to drift freely
until one by one they hit a bounding envelope which grows in size with
$s$.

As $s$ approaches every integer $> 5$, an interesting phenomenon
occurs: A pair of real zeros may meet and become a doubly degenerate
real zero at a point, and then bifurcate into a pair of complex zeros
conjugate to each other. Thus, the pair of real zeros appears to
``collide head-on and scatter perpendicularly'' into a pair of complex
zeros.
\begin{figure}[hbt]
\begin{center}
$\!\!\!\!\!\!\!\!\!\!
\begin{array}{rl}
\raisebox{23pt}{\psfig{figure=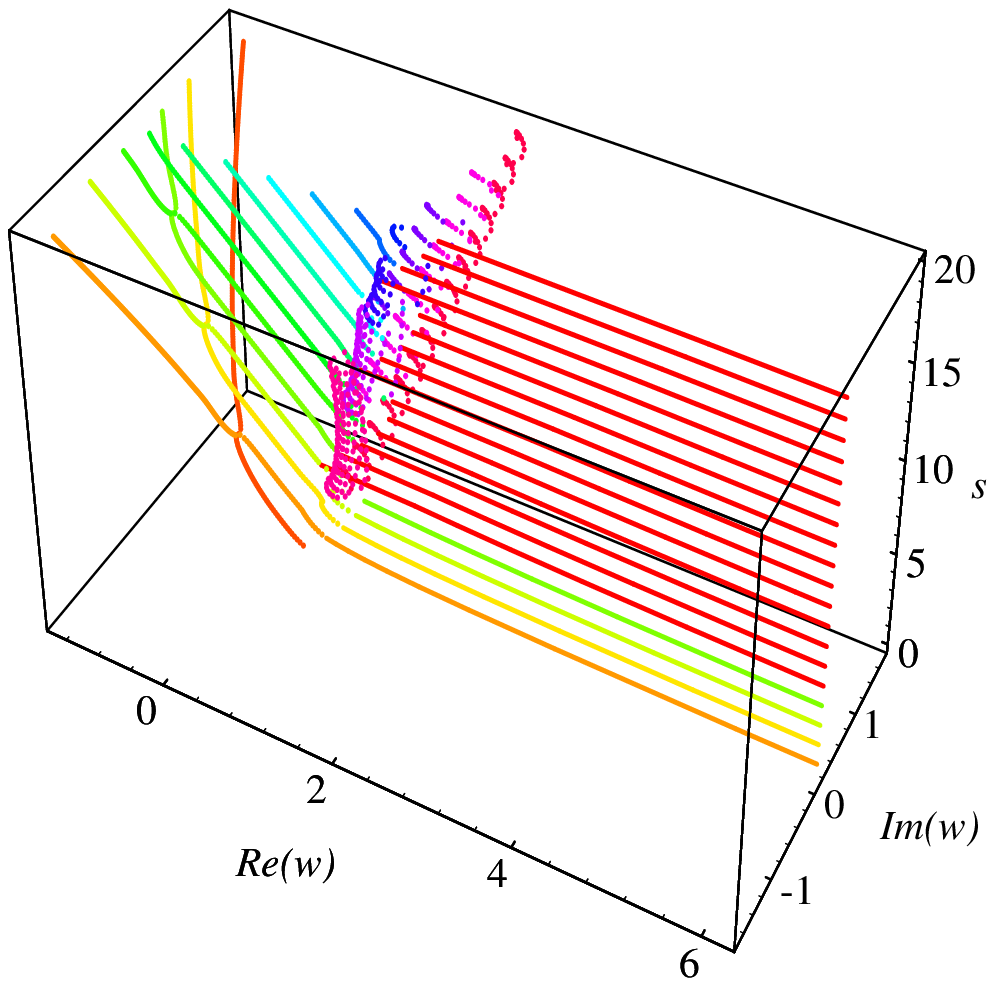,height=162pt}}&
\psfig{figure=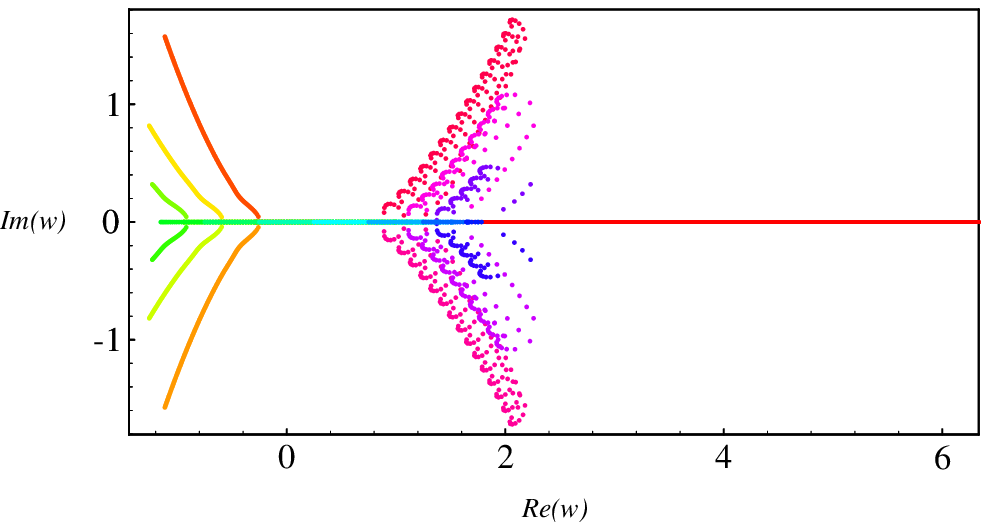,height=208pt}\\
\raisebox{50pt}{\psfig{figure=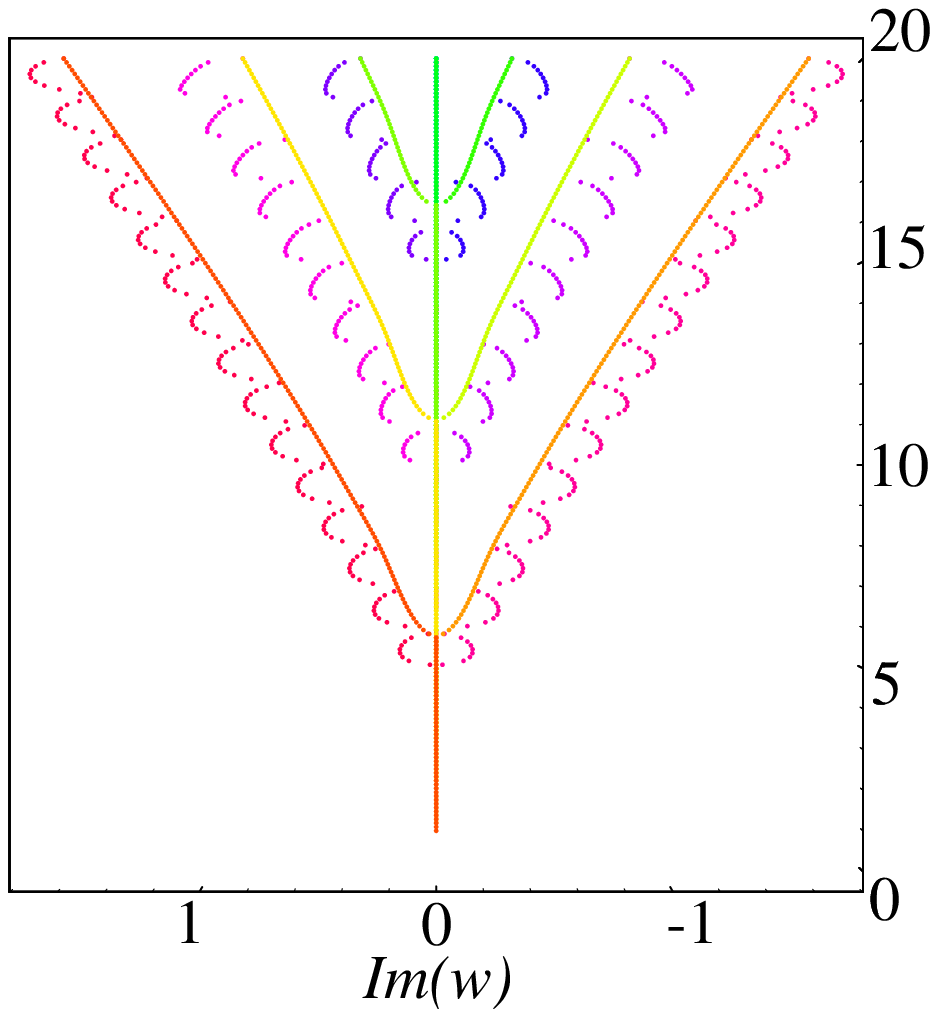,height=103pt}}&
\;\;\;\; \psfig{figure=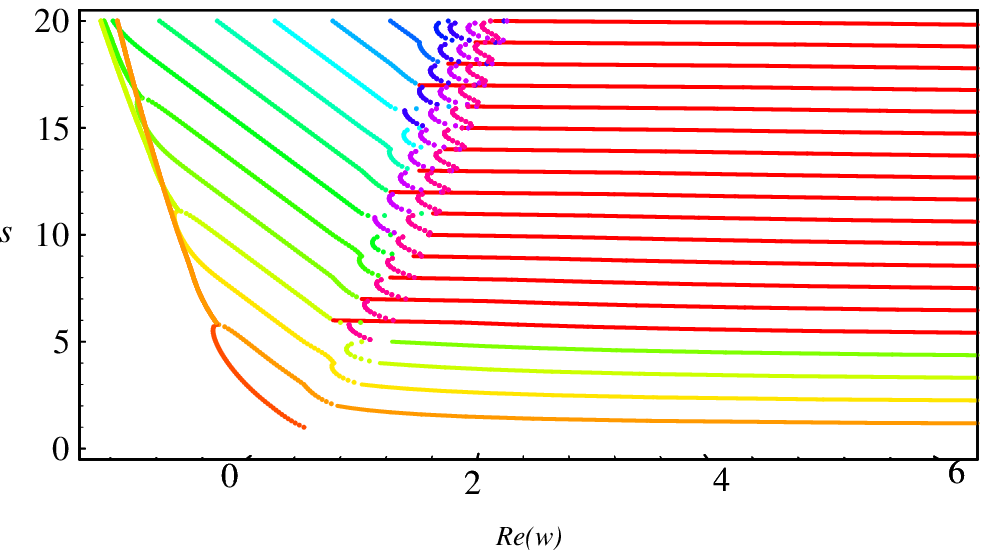,height=198pt}
\end{array}$
\caption{The flow of zeros of $B(s,w)$ forms a complex 3-D structure.}
\end{center}
\end{figure}

\newpage
3 fundamental kinds of scattering can be observed:
\begin{figure}[hbt]
\begin{center}
$\begin{array}{ccc}\!\!\!\!\!\!\!\!\!
\psfig{figure=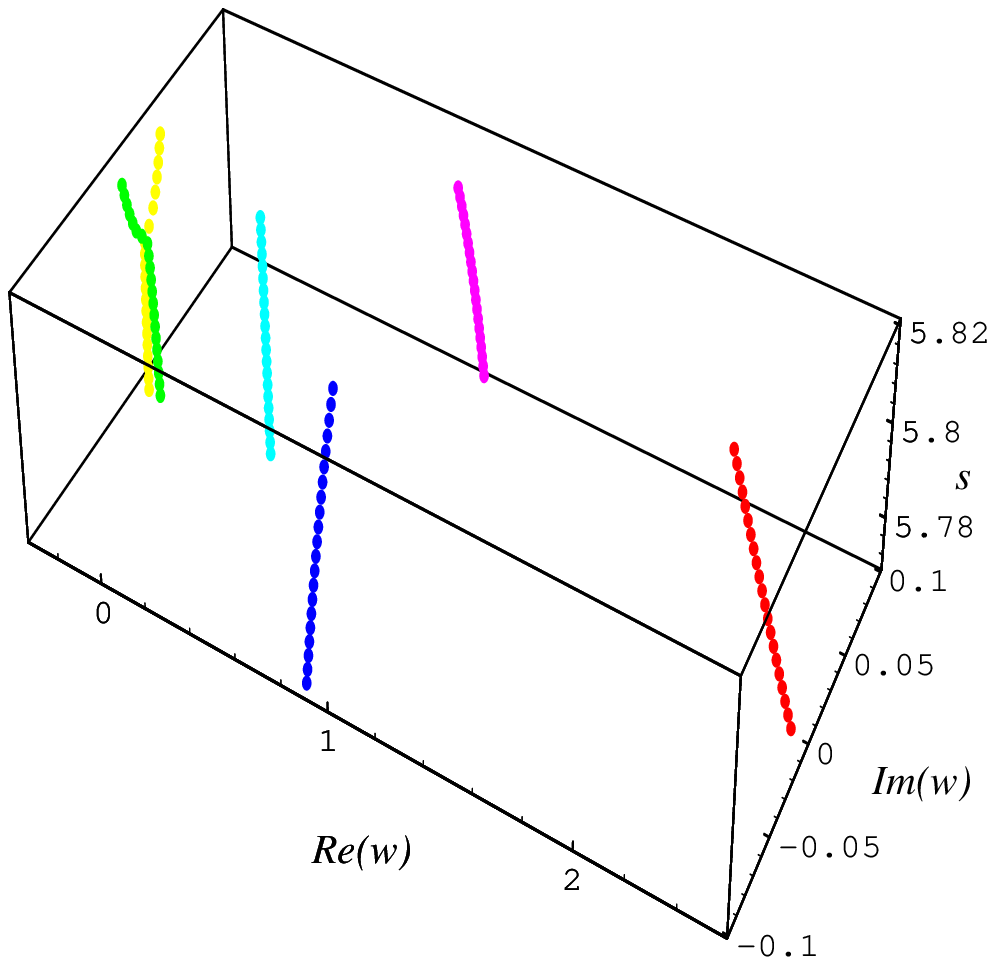,height=128pt}&
\psfig{figure=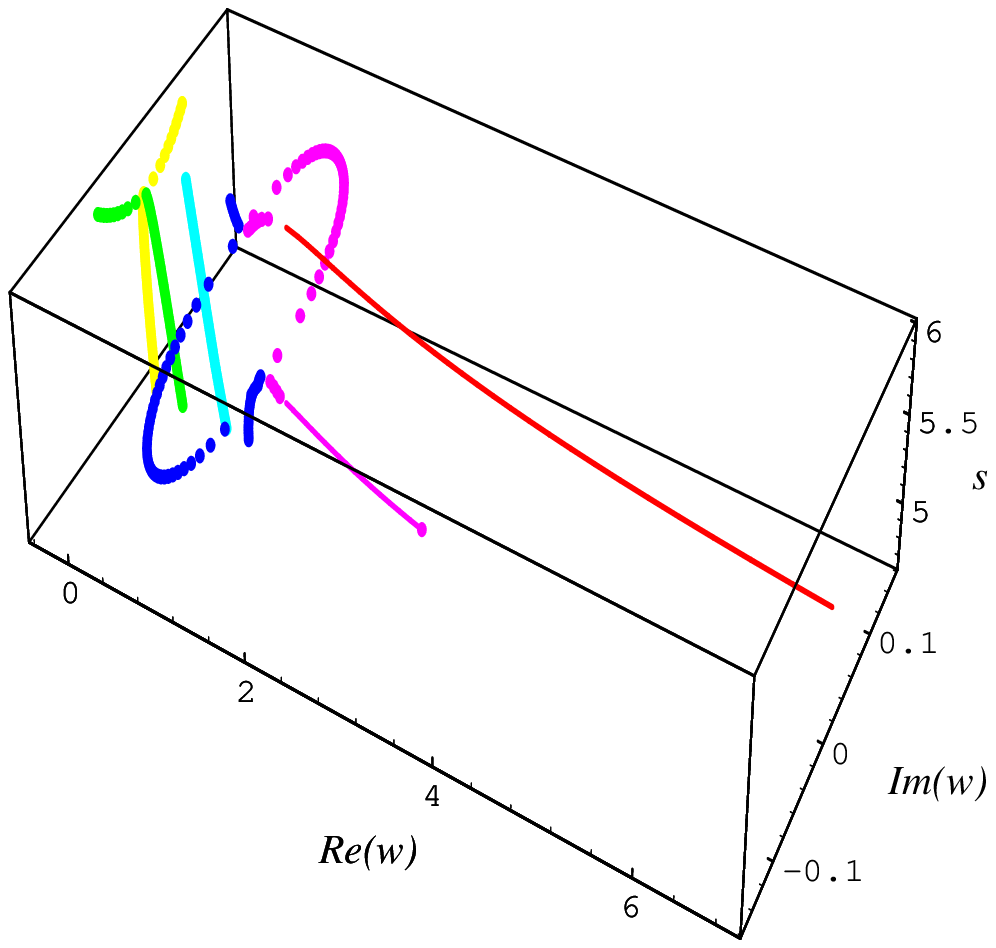,height=128pt}&
\psfig{figure=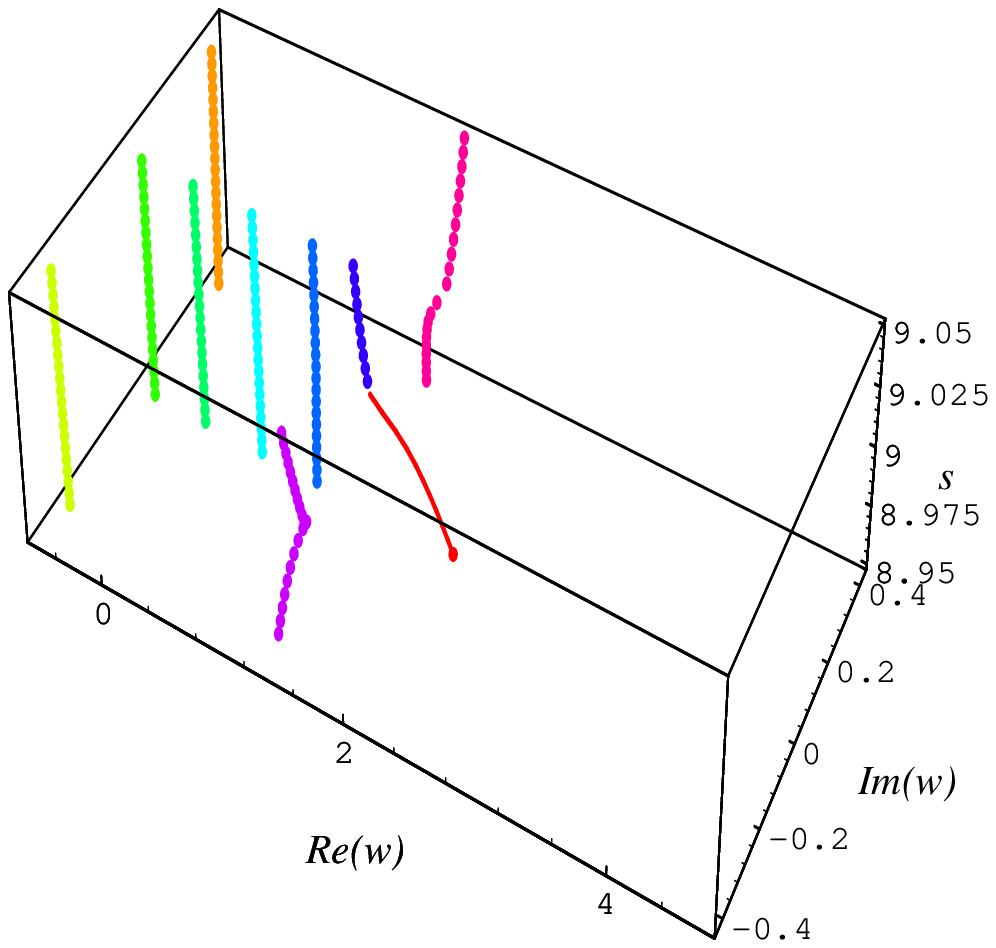,height=128pt}\\
(a) \mbox{ Point} & (b) \mbox{ Loop} & (c) \mbox{ Long-range}
\end{array}$
\caption{3 fundamental kinds of scattering.}
\end{center}
\end{figure}
\begin{itemize}
\item Point scattering:\\
  A pair of real zeros scatter at a point into a pair of complex zeros
  which head away from each other indefinitely.

\item Loop scattering:\\
  The same as point scattering but the pair of complex zeros loops
  back to recombine into degenerate real zeros within unit interval in
  $s$ and then scatter back into a pair of real zeros, much like
  the picture of pair production and annihilation of virtual particles.
  
\item Long-range sideways scattering:\\
  The additional zero that flies in appears as if ``it is carrying
  with it a line front of shockwave'' that stretches parallelly to the
  $\im(w)$ axis. When the ``shockwave'' meets a pair of complex zeros
  that are looping back, the pair gets deflected away from each other
  momentarily before looping back again, while the additional zero
  gets perturbed and slows down discontinuously.
\end{itemize}

The whole complex structure can then be reduced to simple combinations
of these 3 kinds of scattering.
\begin{figure}[hbt]
\begin{center}
$\begin{array}{c}
\psfig{figure=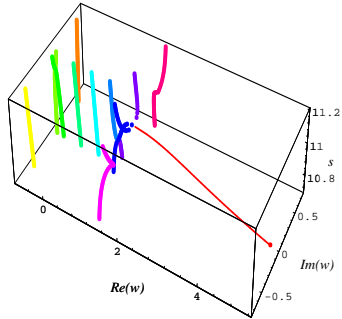,height=128pt}
\end{array}$
\caption{A composite of Point, Loop and Long-range scatterings.}
\end{center}
\end{figure}

The movies (animated gifs) showing the motion of zeros in the complex
$w$ plane at slices of $s$ can be downloaded from Internet at
\cite{woon_movie} and viewed with Netscape or Internet Explorer WWW
browser.

\subsection{Open Challenges in Generalization}
\begin{enumerate}
\item
Generalize the above results of $\,B(s,w)\;$ consistently from
$\,s\in\r\;,\;s \ge 1\;$ to $\,s\in\c\,$.

\item
Derive a set of expressions which give the values of $\,s\,$ where the
point, loop and long-range scatterings occur.
\end{enumerate}

\section{Conclusion}

Bernoulli numbers and polynomials appear in many areas.  In
particular, if we assume the Riemann Hypothesis, when $s\in\,$\c,
$\,$(\ref{zeta(s)_zeta(-n)}) should converge to zero only on the line
$\,s = 1/2 \pm i\/ t$. This remains to be proved. More of this aspect
is analysed in \cite{woon_chaos_order_constants}.

To further self-explore these fascinating properties, feel free to
download and adapt the executable {\em Mathematica} codes from
\cite{woon_codes}. Have fun!

In the meantime, it would be interesting to imagine what Jakob
Bernoulli and Euler would say on these analytic continuations of their
numbers and polynomials.

\vskip .2truein

\noindent {\large \em Acknowledgement}

Special thanks to Y.L. Loh, W. Ballman, B. Lui, P. D'Eath, and K.
Odagiri for discussion, all the friends in Cambridge for
encouragement, and Trinity College and UK Committee of
Vice-Chancellors and Principals (CVCP) for financial support.

\begin{table}[hbt]
{\Large\bf Appendix:}
\begin{center}
{\large \bf Number of real and complex zeros of $B_n(w)$}
\end{center}
$\ba{lll}\!\!\!\!\!\!\!\!\!\!\!\!\!\!
\ba{||c|c|c||}
\hline
B_n(x) & real & complex \\
degree & zeros & zeros \\
n  & \mbox{\bf z}_{\sr(n)} & 
 \mbox{\bf z}_{\sc(n)} \\
\hline\hline
 1 &  1 &  0 \\
 2 &  2 &  0 \\
 3 &  3 &  0 \\
 4 &  4 &  0 \\
 5 &  5 &  0 \\
\hline
 6 &  2 &  4 \\
 7 &  3 &  4 \\
 8 &  4 &  4 \\
 9 &  5 &  4 \\
10 &  6 &  4 \\
11 &  7 &  4 \\
\hline
12 &  4 &  8 \\
13 &  5 &  8 \\
14 &  6 &  8 \\
15 &  7 &  8 \\
16 &  8 &  8 \\
\hline
17 &  5 & 12 \\
18 &  6 & 12 \\
19 &  7 & 12 \\
20 &  8 & 12 \\
21 &  9 & 12 \\
\hline
22 &  6 & 16 \\
23 &  7 & 16 \\
24 &  8 & 16 \\
25 &  9 & 16 \\
26 & 10 & 16 \\
\hline\hline
\ea
\quad
\ba{||c|c|c||}
\hline
B_n(x) & real & complex \\
degree & zeros & zeros \\
n  & \mbox{\bf z}_{\sr(n)} & 
 \mbox{\bf z}_{\sc(n)}\\
\hline\hline
27 &  7 & 20 \\
28 &  8 & 20 \\
29 &  9 & 20 \\
30 & 10 & 20 \\
31 & 11 & 20 \\
32 & 12 & 20 \\
\hline
33 &  9 & 24 \\
34 & 10 & 24 \\
35 & 11 & 24 \\
36 & 12 & 24 \\
37 & 13 & 24 \\
\hline
38 & 10 & 28 \\
39 & 11 & 28 \\
40 & 12 & 28 \\
41 & 13 & 28 \\
42 & 14 & 28 \\
\hline
43 & 11 & 32 \\
44 & 12 & 32 \\
45 & 13 & 32 \\
46 & 14 & 32 \\
47 & 15 & 32 \\
\hline
48 & 12 & 36 \\
49 & 13 & 36 \\
50 & 14 & 36 \\
51 & 15 & 36 \\
52 & 16 & 36 \\
53 & 17 & 36 \\
\hline\hline
\ea
\quad
\ba{||c|c|c||}
\hline
B_n(x) & real & complex \\
degree & zeros & zeros \\
n  & \mbox{\bf z}_{\sr(n)} & 
 \mbox{\bf z}_{\sc}(n)\\
\hline\hline
54 & 14 & 40 \\
55 & 15 & 40 \\
56 & 16 & 40 \\
57 & 17 & 40 \\
58 & 18 & 40 \\
\hline
59 & 15 & 44 \\
60 & 16 & 44 \\
61 & 17 & 44 \\
62 & 18 & 44 \\
63 & 19 & 44 \\
\hline
64 & 16 & 48 \\
65 & 17 & 48 \\
66 & 18 & 48 \\
67 & 19 & 48 \\
68 & 20 & 48 \\
\hline
69 & 17 & 52 \\
70 & 18 & 52 \\
71 & 19 & 52 \\
72 & 20 & 52 \\
73 & 21 & 52 \\
74 & 22 & 52 \\
\hline
75 & 19 & 56 \\
76 & 20 & 56 \\
77 & 21 & 56 \\
78 & 22 & 56 \\
79 & 23 & 56 \\
\hline
80 & 20 & 60 \\
\vdots & \vdots & \vdots \\
\hline\hline
\ea
\ea$
\end{table}

\begin{table}[hbt]
\begin{center}
{\large \bf Solutions of $B_n(w) = 0$}\\
\bigskip
\begin{tabular}{l}
$\ba{||c|l||}
\hline
\bigg. \mbox{ odd } n & 
\quad\quad\quad\quad\quad\quad\quad\quad\quad\quad w\\
\hline\hline
1 & \ba{l}\ds\left(\frac{1}{2}\right)\ea\\
\;&\;\\
3 & \ba{l}\ds\left(\bigg. 0\right),\; \left(\frac{1}{2}\right),\; 
\left(\bigg. 1\right)\ea\\
\;&\;\\
5 & \ba{l}\ds\left(\bigg. 0\right),\; \left(\frac{1}{2}\right),\; 
\left(\bigg. 1\right),\\
\ds\left(\frac{1}{2} - \frac{\sqrt{7/3}}{2}\right)
\approx -0.26376,\; \left(\frac{1}{2} + \frac{\sqrt{7/3}}{2}\right)
\approx 1.26376\ea\\
\;&\;\\
7 & \ba{l}\ds\left(\bigg. 0\right),\; \left(\frac{1}{2}\right),\; 
\left(\bigg. 1\right),\\
\ds\left(\frac{1}{2} - \frac{\sqrt{9 + 2 \sqrt{3} \,i}}{2
    \sqrt{3}}\right) \approx -0.38137 - 0.16376 \,i,\\
\ds\left(\frac{1}{2} - \frac{\sqrt{9 - 2 \sqrt{3} \,i}}{2
    \sqrt{3}}\right) \approx -0.38137 + 0.16376 \,i,\\
\ds\left(\frac{1}{2} + \frac{\sqrt{9 - 2 \sqrt{3} \,i}}{2
    \sqrt{3}}\right) \approx  \;\;\, 1.38137 - 0.16376 \,i,\\
\ds\left(\frac{1}{2} + \frac{\sqrt{9 + 2 \sqrt{3} \,i}}{2
    \sqrt{3}}\right) \approx  \;\;\, 1.38137 + 0.16376 \,i\ea\\
\hline\hline \ea$\\
\quad \\
$\ba{||c|l||}
\hline
\bigg. \mbox{even } n & 
\quad\quad\quad\quad\quad\quad\quad\quad\quad\quad w\\
\hline\hline
2 & \ba{l}\ds
\left(\frac{1}{2} - \frac{1}{2 \sqrt{3}}\right) \approx 0.21133,\;
\left(\frac{1}{2} + \frac{1}{2 \sqrt{3}}\right) \approx 0.78868
\quad\quad\;\,\ea\\
\;&\;\\
4 & \ba{l}
\ds\left(\frac{1}{2} - \frac{\sqrt{15 + 2 \sqrt{30}}}{2
    \sqrt{15}}\right) \approx -0.15770,\\
\ds\left(\frac{1}{2} - \frac{\sqrt{15 - 2 \sqrt{30}}}{2
    \sqrt{15}}\right) \approx \;\;\, 0.24034,\\
\ds\left(\frac{1}{2} + \frac{\sqrt{15 - 2 \sqrt{30}}}{2
    \sqrt{15}}\right) \approx \;\;\, 0.75967,\\
\ds\left(\frac{1}{2} + \frac{\sqrt{15 + 2 \sqrt{30}}}{2
    \sqrt{15}}\right) \approx \;\;\, 1.15770\ea\\
\hline\hline \ea$
\end{tabular}
\end{center}
\end{table}
\end{document}